# Bonding charge distribution analysis of molecule by computation of interatomic charge penetration.

**Yong-U Ri,   Young-Hui Pyon,   Kye-Ryong Sin**[*]

(*Faculty of Chemistry, **Kim Il Sung** University, Pyongyang,*

*Democratic People's Republic of Korea*)

* E-mail: ryongnam9@yahoo.com

**Abstract**

Charge transfer among individual atoms in a molecule is the key concept in the modern electronic theory of chemical bonding. In this work, we defined an atomic region between two atoms by Slater orbital exponents of valence electrons and suggested a method for analytical calculation of charge penetration between all atoms in a molecule. Computation of charge penetration amount is self-consistently performed until each orbital exponent converges to its certain values respectively. Charge penetration matrix was calculated for ethylene and MgO, and bonding charge and its distribution were analyzed by using the charge penetration matrix and the orbital exponents under the bonding state. These results were compared with those by density function method and showed that this method is a simple and direct method to obtain bonding charge distribution of molecule from atomic orbital functions.

**Introduction**

It is significant in investigating characteristics of chemical bond, reactivity of a molecule and interaction between molecules to consider electron distribution in a molecule by atomic region or atomic group region.

Density function theory defined the first-order derivative of total energy of a molecule with respect to the electron density as an chemical potential of electron ( half negative value of electronic negativity of a molecule ), based on its premise that electron density distribution in a molecule exclusively determines the ground state of a molecule.[1] Assuming that electrons will flow between atoms until electronic chemical potential of each atom reaches a certain value, that is, until equalization of them, Rappe and Goddard suggested the charge equalization method to calculate the atomic charge.[2] This is a further development of the orbital electronegativity equalization method[3] proposed by Gasteiger and Morsili. In these methods, the atomic charge and electron transfer were not directly calculated from wave functions. In the case that calculates the atomic charge and electron transfer quantity by electronic distribution function when chemical bond is formed, the way of determining the integration interval, that is, atomic regions in a molecule should be given. Atomic regions in a molecule are defined as being divided by a surface consisted of points in which bonding axis components of electrostatic force equal with each other or by plane perpendicular to bonding axis.[4,5] In the Mulliken analysis[6] defining total charge of the atom by dividing overlap charge of molecular orbital into two terms equally and method of calculating atomic charge using the concept of chemical potential of an electron[7], an atomic region was not specially defined. In the Bader theory for molecules[7], the space partitioning was made by finding the atomic regions in the Euclidean space bounded by a zero-flux surface in the gradient vectors of the one-electron density. From the topological characteristics of the one electron density function there appeared other works dividing a molecular space into the atomic spaces.[8-11] Recently, proposed was a first-principle approach[12] to calculate the charge transfer. Based on the effects of perturbations of an individual atom or a group of atoms on the electron charge density, determined was the amount of electron charge associated with a particular atom or a group of atoms. They computed the topological electron loss versus gain, taking ethylene, graphene, MgO and $SrTiO_3$ as examples.

In this work, we defined atomic region between two atoms by Slater orbital exponents of valence electrons and suggested a method for analytical calculation of charge penetration between all atoms in a molecule by using the one-electron density function of atoms. Charge penetration was self-consistently calculated until the each orbital exponent converged to its certain value respectively. Charge penetration matrix was calculated for ethylene and MgO, and bonding charges for each atom and bonding charge distributions in molecules are analyzed by using the charge penetration matrix and the orbital exponents under bonding state.

## Theory and computation method

In this work, the molecular bonding charge distribution was made up of only the change of distribution of valence electrons due to the formation of chemical bonds. This allows the simple calculation of the bonding charge distribution in macromolecules and solids without heavy cost of calculation.

### Computation of charge penetration matrix

From the view of one electron approach, a molecular space can be partitioned into atomic regions according to magnitudes of force of atomic centers. In two-center field consisted of two atoms, the surface composed of the points at which an electron was influenced by equal electrostatic force from two atoms was defined as the boundary surface partitioning molecular space into two atomic regions.

The effective charges of two nuclei (r, s) were denoted as $z_r^*$, $z_s^*$ and the distance from an electron at the given point to the nuclei as $r_r$ and $r_s$, and the orbital exponents and effective principal quantum numbers of two Slater orbitals forming chemical bond as $\zeta_r$, $n_r^*$, $\zeta_s$, $n_s^*$ respectively.

If electrostatic forces of two nuclei exerted on the given electron are equal, the following relations will be established.

$$\frac{z_r^*}{r_r^2} = \frac{z_s^*}{r_s^2}$$

$$\frac{r_r}{r_s} = \left(\frac{z_r^*}{z_s^*}\right)^{1/2} = \left(\frac{\varsigma_r}{\varsigma_s}\right)^{1/2}\left(\frac{n_r^*}{n_s^*}\right)^{1/2} \equiv \gamma \tag{1}$$

As shown in Eq.(1), in two center field, two atomic regions were defined by the surface composed of points at which the ratio of distances ($r_r$, $r_s$) from two nuclei is $\gamma$.

Then the equation of the surface was given as follws:

$$x^2 + y^2 + (z - \frac{\gamma^2}{\gamma^2 - 1}R)^2 = (\frac{\gamma R}{\gamma^2 - 1})^2 \tag{2}$$

where R is a distance between two atoms and z-axis is coincided with bond axis. If $\gamma \neq 1$, the surface will be a sphere laid on Z axis and as $\gamma$ comes to 1 the sphere approaches a plane passing bond center.

For atomic pair r-s, as the boundary surface described in Eq.(2) was given, electron charge penetration from each atom to the other atomic region can be analytically calculated.

Charge penetration of r atom to s atom, in other words, the probability $D_{rs}$ that valence electron from atom ( r ) appears in the region of atom (s) can be defined in cylindrical c oordinates as follows:

$$D_{rs} = \int_0^{2\pi} d\varphi \int_{z_1}^{z_2} dz \int_0^L \psi_r^2(\rho, z, \varphi) \rho d\rho \tag{3}$$

where intervals of integration $z_1$, $z_2$, L were analytically determined by Eq.(2) like the following:

$$L = [(r_0^2 - (z-a)^2]^{1/2}$$

$$z_1 = \gamma R/(1+\gamma)$$
$$z_2 = \gamma R/(\gamma - 1)$$
$$a = \gamma^2 R/(\gamma^2 - 1)$$
$$r_0 = \gamma R/(\gamma^2 - 1)$$

Here $\psi$ is the Slater atomic orbital and described in the atomic units as follows:

$$\psi = N_n r^{n^*-1} \exp(-\varsigma r) \cdot Y_{lm}(\theta, \varphi) \tag{4}$$

$$N_n = [(2\varsigma)^{2n+1}/(2n)!]^{1/2}$$

where $Y_{lm}(\theta, \phi)$ is the directional function and $N_n$ is the normalizing constant. From Eq.(3) $D_{rs}$ depends on atomic region, kind of atomic orbital and interatomic distance. And $\zeta_i$, Slater orbital exponent of the i-th electron, can be determined by the following equation.[13]

$$IP_i = R_0 \varsigma_i^2 \tag{5}$$

where $R_0$=13.6eV and the orbital ionization potential IP was determined by the following equation (6). [13]

$$IP_i(n,l,m) = A_i(n,l,m)Q^2 + B_i(n,l,m)Q + C_i(n,l,m) \qquad (6)$$

where Q is atomic charge, $A_i$, $B_i$ and $C_i$ are constants determined by electronic configuration( *n, l, m*) in the given isoelectronic series.

In two-centered bond the z-axis of atomic fixed coordinate system can be coincided with that of bonding coordinate system. In this case, the charge penetration is called the standard charge penetration and denoted as $D_{rs}^s$ by using a superscript 's'.

For polyatomic molecule, the atomic fixed coordinate system can be set up to be parallel with the molecular coordinate system, but that atomic fixed coordinate is not generally coincided with the bonding coordinate. In this case, the atomic orbital denoted in the atomic fixed coordinate system can be decomposed into the atomic orbitals denoted in bonding coordinate system for simple calculation of the charge penetration. The calculation of charge penetration follows the same way as calculating the generalized overlap integral by using the standard overlap integral but the difference is that elements of transformation matrix in charge penetration calculation are expressed as square of elements of transformation matrix in overlap integral.

For $p_z$, $p_x$ and $p_y$ orbitals, the standard charge penetration $D_p^s$ in bonding coordinate system is changed into charge penetration $D_p$ in atomic fixed coordinate system by the following transformation matrix:

$$\begin{pmatrix} D_{p_z} \\ D_{p_x} \\ D_{p_y} \end{pmatrix} = \begin{pmatrix} \cos^2\alpha & 0 & \sin^2\alpha \\ \sin^2\alpha\cos^2\beta & \sin^2\beta & \cos^2\alpha\cos^2\beta \\ \sin^2\alpha\sin^2\beta & \cos^2\beta & \cos^2\alpha\sin^2\beta \end{pmatrix} \begin{pmatrix} D_{p_z}^s \\ D_{p_x}^s \\ D_{p_y}^s \end{pmatrix} \qquad (7)$$

The same holds true for charge penetrations of d, f orbitals.

In polyatomic molecules where a valence electron moves around several atomic cores, the charge penetration of a given electron should be calculated considering the interactive effects by the other atoms.

For an atom with several valence electrons, it is useful to introduce an average orbital of atom for simple calculation of charge penetration.

The average orbital of atom introduced here is a kind of Slater Type Orbital (STO), where its radial part is determined by its orbital exponent ($\zeta$) that can be calculated from mean ionization energy of valence electrons by the following equation (8):

$$\varsigma = \left[ \frac{n_s IP_s + n_p IP_p + n_d IP_d}{n_s + n_p + n_d} / R_0 \right]^{1/2} \qquad (8)$$

The charge penetration calculated by this average atomic orbital can be considered as a charge penetration value between each atomic pair in poly atomic molecule and let's denote it by using a superscript '0' like $D_{rs}^0$ for the charge penetration for r-s atomic pair.

An electron in r-atom can be shared with other atoms according to the following partition function $F_r$:

$$F_r = \frac{1}{1 + \sum_s D_{rs}^0/(1 - D_{rs}^0)} \qquad (9)$$

Therefore, an electron in r-atom is shared with arbitrary s-atom in the molecule to the following extent:

$$D_{rs} = \frac{D_{rs}^0}{1 - D_{rs}^0} F_r \qquad (10)$$

The charge penetration matrix TD of the molecule is calculated as following:

$$TD_{rs} = D_{rs} \cdot N_r$$
$$TD_{sr} = D_{sr} \cdot N_s$$

where $N_r$, $N_s$ are numbers of valence electron of r-, s-atoms respectively.

The charge penetration of adjacent atoms leads to the change of valence orbital exponent of the given atom.

$$\varsigma_r = \varsigma_r^0 + \frac{\sigma}{n^*} \sum_s TD_{sr} \qquad (11)$$

where $\varsigma_r^0$ is a atomic orbital exponent in the free atom and $\sigma$ is the Slater screening constant.

The orbital exponent is continuously changed with the change of interatomic charge penetration by Eq.(11) until it reaches a certain equilibrium values respectively according to principle of the charge equalization[2], when the equilibrium electron distribution corresponding to the stable molecular structure can be considered to be achieved.

If charge penetration matrix TD of the molecule was calculated, the atomic charge of each atom would be calculated as follows:

$$q_r = \sum_s (TD_{rs} - TD_{sr}) \qquad (12)$$

**Computation of bonding charge distribution of molecule**

The previous researchers[12] defined the bonding charge density as the electron transfer among the atoms relative to the non-bonded free atoms when forming the chemical bond as follows.

$$\Delta \rho_S(r) = \rho_S(r) - \rho_{S0}(r) \qquad (13)$$

where $\rho_{s0}(r)$ represents the non-interacting electron charge density calculated from a superposition of electron charge densities of the free atoms.

Based on the charge penetration matrix of molecule and the orbital exponents when the orbital exponents reaches a certain equilibrium values respectively, we calculated the bond charge density and the bond charge density distribution in molecule.

The diagonal elements in charge penetration matrix TD of molecule are numbers of the remained electrons in the given atoms and the non-diagonal elements are numbers of the transferred electrons from the given atom to another atom. And the sum of each column elements is number of the electrons partitioned to each atom in the bonding state. Therefore, from the difference between TDs in the bonding and nonbonding state, we can find out the bonding electrons distribution among atoms or atom groups.

For calculating the bond charge density distribution, let's denote the average one-electronic density function of each atom as follows.

$$\rho_r = \rho(x - x_r, y - y_r, z - z_r, \varsigma_r)$$

where $x_r$, $y_r$, $z_r$ are the coordinates of r-atom and $\zeta_r$ are the average atomic orbital exponents of r-atom in the nonbonding state.

Postulating electrons penetrated to each atom followed the electron density distribution of that atom, total electron density distribution in the nonbonding and bonding states, $\rho_{0M}$ and $\rho_M$, can be expressed respectively as follows.

$$\rho_{0M} = \sum_{r=1}^{N} n_r \rho_r \tag{14}$$

$$\rho_M = \sum_{s=1}^{N}\sum_{r=1}^{N} TD_{rs}\rho_s + 2\sum_{s<t}^{N}\sum_{r=1}^{N}[TD_{rs}TD_{rt}\rho_s\rho_t]^{1/2} \tag{15}$$

where N is number of total atoms and $n_r$ is number of valence electrons in the r-atom.

Bonding electron density distribution of molecule is calculated as follows.

$$\Delta\rho_M = \rho_M - \rho_{0M} \tag{16}$$

Total charge penetration amount in molecule can be obtained through self-consistent calculation with Eq.(11) until each orbital exponent converges at its certain value respectively.

## Computation and discussion

Here ethylene and MgO were chosen as an example for calculating the bonding charge distribution in a molecule or solid by using the above-mentioned charge penetration matrix and Eq.(11).

Entry data necessary in calculation of charge penetration matrix of a molecule include the geometric structure of the molecule, orbital ionization potentials of valence electrons and parameter $\sigma$ in Eq.(15). The numeric value of $\sigma$ was chosen as 0.3 for 1s orbital and as 0.35 for (ns,np) orbitals.

Orbital ionization potential was determined as follows. The orbital ionization potentials for the elements in the 1st~3rd periods of the periodic system and in the first of the transition metal series were calculated by using the electronic configuration in isoelectronic sequence by H.Basch[13].

The orbital ionization potentials of some atoms used as entry date were given in Table 1.

Table 1. Electron number and orbital ionization potentials IP(eV) of valence electrons of atom.

| atom | Atom number | principal quantum number | electron number and ionization potential(eV) | | | | | |
|---|---|---|---|---|---|---|---|---|
| | | | $n_s$ | $IP_s$ | $n_p$ | $IP_p$ | $n_d$ | $IP_d$ |
| H | 1 | 1 | 1 | 13.600 | | | | |
| Li | 3 | 2 | 1 | 3.397 | | | | |
| Be | 4 | 2 | 1 | 9.320 | 1 | 5.892 | | |
| B | 5 | 2 | 1 | 15.156 | 2 | 8.306 | | |
| C | 6 | 2 | 1 | 19.440 | 3 | 10.670 | | |
| N | 7 | 2 | 2 | 25.580 | 3 | 13.190 | | |
| O | 8 | 2 | 2 | 32.333 | 4 | 15.795 | | |
| F | 9 | 2 | 0 | 40.229 | 1 | 18.646 | | |
| Na | 11 | 3 | 2 | 80.110 | 6 | 40.400 | | |
| Mg | 12 | 3 | 1 | 7.640 | 1 | 4.520 | | 1.740 |
| Al | 13 | 3 | 1 | 11.320 | 2 | 5.970 | | 1.960 |
| Si | 14 | 3 | 1 | 15.150 | 3 | 5.620 | | 2.040 |
| P | 15 | 3 | 2 | 19.370 | 3 | 10.840 | | 2.960 |
| S | 16 | 3 | 1 | 20.520 | 3 | 10.780 | | 3.060 |
| Cl | 17 | 3 | 2 | 25.290 | 5 | 13.990 | | 1.000 |
| K | 19 | 3 | 2 | 48.030 | 6 | 31.920 | | |
| Ca | 20 | 3 | 2 | 70.140 | 6 | 51.340 | | |

The atomic coordinates of ethylene were determined from its well-known geometric structure [14] and MgO unit cell used in computation was taken from the experimental measurement of its crystal structure[15] and the computation was carried out on the cubic structure model composed of 5×5×5 = 125 atoms.

For ethylene molecule and MgO lattice, the bonding electron density of each atom was calculated from the charge penetration matrix in bonding state and results were compared with literature data[12] obtained by DFT calculations.(Table 2)

Table 2. Total number of electrons and bonding charges in atoms

| Atom | C | H | Mg | O |
|---|---|---|---|---|
| total number of electrons in initial state | 3.864 | 1.068 | 1.422 | 6.584 |
| total number of electrons in bonding state | 4.155 | 0.923 | 1.742 | 6.260 |
| bonding charges | 0.155 | -0.077 | -0.250 | 0.260 |
| bonding charges[12] | 0.289 | -0.145 | -0.531 | 0.527 |

In Table 2, the literature data[12] are the total effective bonding charges obtained by integrating the charge redistribution of the perturbed (infinitesimally displaced) atoms in some space. Total number of electrons in initial state was calculated from the initial charge penetration matrix and that in bonding state was from the equalized charge penetration matrix. And bonding charges are the difference of total number of electrons between the bonding and nonbonding states.

As shown in Table 2, the bonding charges calculated from charge penetration matrix are smaller than those calculated from atomic perturbation by DFT[12]. But two results have similar tendency with respect to electron transfer by the formation of chemical bond.

The total charge density distributions in nonbonding and bonding state were calculated by Eq.(18) and Eq.(19), and the bonding charge density distributions were calculated by Eq.(20).

Figure 1 shows the bonding charge distribution in the molecular plane of ethylene, where the charge density increases according to the color sequence of orange-yellow-green-blue-red.

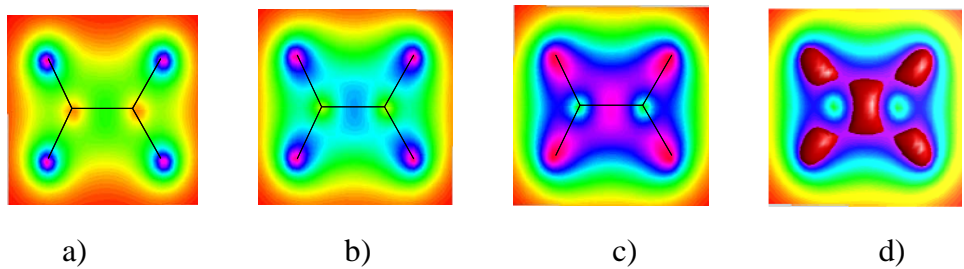

      a)            b)            c)            d)

Fig.1. Total charge density distributions of nonbonding state(a), bonding state(b) and bonding charge density distribution(c, d) in the molecular plane of ethylene.

In Figure 1, part (a) is total charge density distribution calculated by Eq. (18) using initial orbital exponents and part (b) is that calculated by Eq. (19) after charge equalization. And part (c) is bonding charge distribution calculated by Eq. (16) and part (d) is the isosurface of bonding charge density with charge of 0.21. In the equalized state of charge penetration, the bonding electrons are concentrated in the middle region of molecule and the charge density values are particularly large between two carbons. This illustrates well the covalent bond character of C-C bond in ethylene.

Similarly the bonding charge distributions were calculated in MgO lattice.

In Figure 2, part (a) and (b) are the images of total charge distribution in nonbonding and bonding states and part (c) is the image of bonding charge distribution in the middle layer of MgO lattice. Part (d) is the contour image of bonding charge distribution. In the image (a) and (b), the red circlets are oxygen atoms and Mg atoms are placed between them.

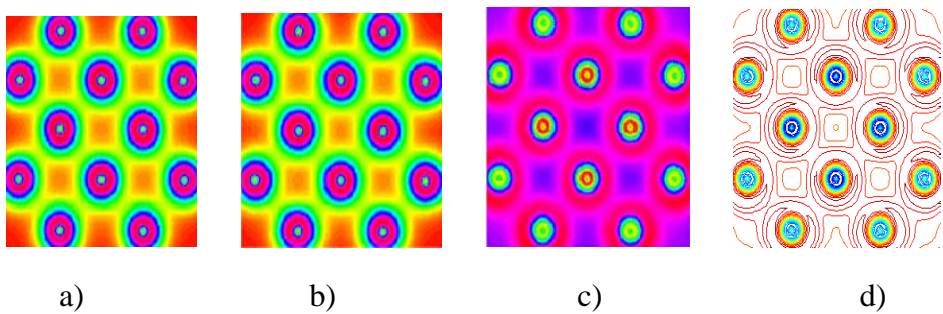

      a)            b)            c)            d)

Fig. 2. Total charge density distributions of nonbonding state(a), bonding state(b) and bonding charge density distribution(c, d) in (001) plane of MgO lattice.

As shown in Fig.2, unlike ethylene, in MgO the change of charge density distribution in nonbonding and bonding states was not large. The image (c) of the bonding charge distribution shows that bonding charge values are positive only in the region surrounding oxygen atoms but negative in the region near magnesium atoms, and are nearly zero in the middle region between atoms. This indicated that Mg-O bond has ionic bond character, which is well compared with the experimental evidence that MgO was a crystal with pronounced ionic character[12].

## Conclusions

In this paper, presented was a new definition of atomic region among two atoms by Slater orbital exponents of valence electrons and suggested a method for analytical calculation of charge penetration between all atoms in a molecule or crystal. Computation of charge penetration amount was self-consistently performed until each orbital exponent converges at its certain values respectively.

For ethylene and MgO crystal as examples, charge penetration matrix was calculated and bonding charges for each atom and bonding charge distributions in molecules were analyzed by using the charge penetration matrix and the orbital exponents under bonding state. The results were compared with the previous DFT calculations and showed that this proposed method can be considered as a simple method to obtain bonding charge distribution of molecule directly from STO atomic orbitals without solving Hartree-Fock ( or Kohn-Sham ) equation over the whole molecule.